\documentclass[sigconf]{acmart}

\AtBeginDocument{%
  \providecommand\BibTeX{{%
    \normalfont B\kern-0.5em{\scshape i\kern-0.25em b}\kern-0.8em\TeX}}}

\copyrightyear{2024}
\acmYear{2024}
\setcopyright{acmlicensed}
\acmConference[KDD '24] {Proceedings of the 30th ACM SIGKDD Conference on Knowledge Discovery and Data Mining }{August 25--29, 2024}{Barcelona, Spain.}
\acmBooktitle{Proceedings of the 30th ACM SIGKDD Conference on Knowledge Discovery and Data Mining (KDD '24), August 25--29, 2024, Barcelona, Spain}
\acmISBN{979-8-4007-0490-1/24/08}
\acmDOI{10.1145/3637528.3671786}

\usepackage{algorithm}
\usepackage{enumitem}
\usepackage{graphicx}
\usepackage{subcaption}
\usepackage{mathrsfs} 
\usepackage{bm}
\usepackage{bbm}
\usepackage{microtype}
\usepackage{algpseudocode}
\usepackage{amsmath}
\usepackage{booktabs}

\begin{document}

\title{Performative Debias with Fair-exposure Optimization Driven by \\ Strategic Agents in Recommender Systems}


\author{Zhichen Xiang}
\authornote{This work was done when Zhichen Xiang was doing an internship at AI Lab of Lenovo Research.}
\affiliation{%
  \institution{College of Management and Economics \& Laboratory
of Computation and Analytics of Complex Management Systems (CACMS), Tianjin University }
  \city{Tianjin}
  \country{China}}
  \email{xiangzc@tju.edu.cn}

\author{Hongke Zhao}
\authornote{Corresponding Authors}
\affiliation{%
  \institution{College of Management and Economics \& Laboratory
of Computation and Analytics of Complex Management Systems (CACMS), Tianjin University }
  \city{Tianjin}
  \country{China}}
  \email{hongke@tju.edu.cn}

\author{Chuang Zhao}
\affiliation{%
  \institution{The Hong Kong University of Science and Technology }
  \city{Hong Kong}
  \country{China}}
  \email{czhaobo@connect.ust.hk}

\author{Ming He}
\authornotemark[2]
\affiliation{%
 \institution{AI Lab at Lenovo Research }
 \city{Beijing}
 \country{China}}
 \email{heming01@foxmail.com}

\author{Jianping Fan}
\affiliation{%
 \institution{AI Lab at Lenovo Research }
 \city{Beijing}
 \country{China}}
 \email{jfan1@lenovo.com}
\renewcommand{\shortauthors}{Zhichen Xiang, Hongke Zhao, Chuang Zhao, Ming He, \& Jianping Fan}

\begin{abstract}
Data bias, e.g., popularity impairs the dynamics of two-sided markets within recommender systems. This overshadows the less visible but potentially intriguing long-tail items that could capture user interest. Despite the abundance of research surrounding this issue, it still poses challenges and remains a hot topic in academic circles. Along this line, in this paper, we developed a re-ranking approach in dynamic settings with fair-exposure optimization driven by strategic agents. Designed for the producer side, the execution of agents assumes content creators can modify item features based on strategic incentives to maximize their exposure. This iterative process entails an end-to-end optimization, employing differentiable ranking operators that simultaneously target accuracy and fairness. Joint objectives ensure the performance of recommendations while enhancing the visibility of tail items. We also leveraged the performativity nature of predictions to illustrate how strategic learning influences content creators to shift towards fairness efficiently, thereby incentivizing features of tail items. Through comprehensive experiments on both public and industrial datasets, we have substantiated the effectiveness and dominance of the proposed method especially on unveiling the potential of tail items.

\end{abstract}




\maketitle

\section{Introduction}

Recommender systems play a pivotal role in curating high-quality, personalized content in two-sided markets, encompassing both customers and producers~\cite{naghiaei2022cpfair,patro2020fairrec}. Algorithms are designed to bridge the gap between consumer desires and the vast array of available products. Considering sustainability, it is critical to build a fairer two-sided platform that understands and optimizes the dynamics of human-algorithm interactions~\cite{brinkmann2023machine,piao2023human,zhou2023longitudinal,zhu2021popularity}, as well as identifying and alleviating bias in dynamic systems with human feedback loops~\cite{jiang2019degenerate,mansoury2020feedback,sinha2016deconvolving,sun2019debiasing}. However, it's known that popularity bias would lead to unfair exposure for tail items\cite{chen2023bias,wei2021model,zhao2023ensemble}. This bias occurs as popular items receive disproportionate recommendations, regardless of their quality. 
That causes a self-reinforcing cycle exacerbating data imbalance in the future, leading to the ``Matthew effect'', where the rich get richer and the poor get poorer\cite{zhu2021popularity}.

\begin{figure}[t]
  \centering
  \includegraphics[width=\linewidth]{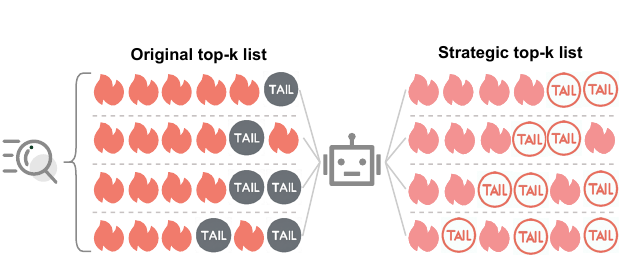}
  \caption{Function of the strategic agent: improve the performance of tail items in top-k recommendation via feature incentives from users' candidate lists.}
  \label{concept figure}
\end{figure}

The prevalent research concerning re-ranking for addressing popularity bias involves implementing post-hoc algorithms to adjust the outputs of conventionally trained ranking models, thereby shifting focus towards tail items. This includes approaches such as enhancing the visibility of less popular items~\cite{abdollahpouri2019managing,abdollahpouri2021user}, and employing diversification techniques, such as traversing the list of ranked items to eliminate those that are similar to items ranked higher~\cite{carbonell1998use,sha2016framework}. However, recommendations are inherently dynamic, and these existing post-hoc approaches may not adequately improve the intrinsic utility of long-tail items over the long term. Furthermore, they overlook the operational reality of recommendation systems within two-sided markets. Moreover, consumer fairness is mainly about addressing disparities within user groups based on sensitive attributes on platforms~\cite{kotary2022end,mahapatra2023multi,morik2020controlling,zehlike2020reducing,chen2023improving}. In contrast, our study is dedicated to alleviating the disparities in item exposure level that arise due to systemic popularity biases. Encouraging the producer to make efforts, we aim to unearth the beneficial features of tail items during dynamics in two-sided markets.

Based on this, we have designed a ranking policy updating based on the strategy learning agent\cite{hardt2016strategic} executed by producers to address the disadvantages of tail items. In order to formalize the endeavor of producers in two-sided markets, we assume that items on the platform are shaped by content creators~\cite{ben2020content,hron2022modeling}. The goal of each creator is to maximize their own exposure. In addition, it can modify item characteristics based on incentives of strategy agents to achieve maximum utility that meets its audience. In order to better utilize the content of items, we extract as many semantic features as possible and focus the ranking algorithm on the inner product of user representations and item features. Figure \ref{concept figure} shows the intuitive role of the strategic agent.

To guide content creators in updating, we propose dual optimization objectives, that ensure both accuracy and fair exposure in top-$k$ lists for users. Recognizing the challenge presented by user profiles shifting towards popular items~\cite{chaney2018algorithmic}, which may lead to content homogenization, we seek to balance two factors effectively. However, the non-differentiable nature of these metrics makes gradient descent algorithms unsuitable for joint optimization. To solve this challenge, differentiable ranking operators are designed to integrate them into the process of end-to-end training, and our goal is to explore their potential relationship and suggest methods for optimally balancing this trade-off.

Upon completing the development of the differentiable operators, we investigate the system's effective utilization of evolution power by leveraging performative prediction~\cite{perdomo2020performative}. This concept encompasses strategic learning in a temporal setting, where ongoing retraining and learning lead to shifts in the underlying data over time. Specifically, by implementing the execution rule of strategic agents, we enhance the fair-exposure through anticipative regularization. Updated representations are employed to calculate the fairness metrics for the forthcoming epoch, which informs the optimization process in the current epoch. Consequently, that enables the model to benefit from more effective guidance of achieving optimization oriented towards fairness.

In each round of optimization, the objective focused on fair exposure enables the user representations trained in the current time to include more information about tail items (those with lower relevance) of the candidate list. Responding to the evolving representations of users, content creators strategically modify item features to meet the optimal utility for their target users in the following iteration. These multi-round iterative optimizations enable the strategic agent to execute multiple times and fully incentivize the features of tail items in top-$k$ lists. This strategic learning mechanism capitalizes on the performativity of the anticipated user representations, to infuse relevant feature incentives concerning fairness into items across various popularity levels for debias efficiently. By doing so, producers are able to identify and leverage the beneficial features of tail items, thereby fostering the sustainable growth of two-sided markets in recommendation.

In summary, this article makes the following contributions:
\begin{itemize}[leftmargin=*]
    \item This paper addresses the issue of popularity bias in recommender systems from the perspective of producers' strategic behavior within two-sided markets during the re-ranking stage. We leverage the performativity of prediction towards fair-exposure optimization for content creators, which efficiently mines the potential of long-tail items for users.
    \item We have developed a differentiable operator based on the fairness of item exposure. This operator allows for end-to-end dual-objective optimization, balancing both the accuracy and fairness of the recommendation results.
    \item Extensive experiments have been conducted on both public and industrial datasets. The results validate the effectiveness and rationality of the proposed methods.
\end{itemize}

\section{Related Work}
The two most relevant research areas of this paper are popularity bias in recommendation and performative prediction. We introduce the related research work from these two aspects respectively.

\subsection{Popularity Bias in Recommendation}

At present, the relevant research on popularity bias in recommendation systems can be divided into two categories, one is static data debias, and the other is debias in dynamic scenarios~\cite{chen2023bias}.

From the perspective of the static data, classical methods to deal with popularity bias are using the Inverse Propensity Score (IPS) ~\cite{li2023propensity,schnabel2016recommendations} and casual model~\cite{gao2023cirs,wei2021model,zhang2021causal,zheng2021disentangling,zhao2023sequential}. Besides, the learning-to-rank algorithm based on regularization~\cite{abdollahpouri2017controlling} is also introduced to enhance the long-tail coverage of recommendation lists. For specific scenarios, \cite{cheng2024general}utilized the contextual information of tail items to enhance their representation in sequential recommendations, thereby promoting their performance.

As for the dynamic scenario, the behavior of user engagement is a valuable signal for both the user and the platform where the algorithm is deployed~\cite{brinkmann2023machine,morik2020controlling,sun2019debiasing}, enabling them both to be connected into complex feedback loops. \cite{chaney2018algorithmic}analyzed the feedback loop between user behavior and algorithmic recommendation system, and revealed how algorithm-based confounding factors in recommendation system increase the homogeneity of user behavior and reduce the utility. Empirical research is also carried out through simulation experiments~\cite{zhu2021popularity}, and the inherent imbalance of audience size and model bias in dynamic scenes are analyzed as the main sources of popularity bias. Overall, the current research about dynamic debias predominantly focuses on simulating human behavior, with scant attention given to analyzing the endeavors of producers.

\subsection{Strategic Learning and Performativity}
Strategic learning~\cite{hardt2016strategic} requires the model to be able to identify and adapt to participants' strategic behavioral changes, thus creating a feedback loop. These forecasts are notably characterized by the model's adaptability to alterations in the environment instigated by its own predictions. 
In this context, forward-looking regularization~\cite{rosenfeld2020predictions} and causal identifiability~\cite{mendler2022anticipating} are proposed, encouraging predictive models to induce changes that also improve outcomes by predicting user actions.

Performative prediction~\cite{perdomo2020performative} refers to the fact that in the prediction process, the prediction result itself will affect the occurrence of the predicted event, and the output of the prediction model will, in turn, affect its input data distribution. This is called ``predicting from predictions''. Existing studies provide sufficient global conditions for retraining convergence~\cite{perdomo2020performative} or propose general optimization algorithms~\cite{brown2022performative,miller2021outside}, which are set up in a way that allows complex human interactions in loops~\cite{taori2023data}. The performative recommendation is firstly proposed to diversify content creators~\cite{eilat2023performative}. However, it lacked evidence from real-world data and paid less attention to the exposure of items with different popularity. In addition, some studies~\cite{liu2022strategic,jagadeesan2024supply,hron2022modeling} also introduce strategy adaptation into systems built by content creators, which capture competition between
participants and formally exploring the strategic behavior of producers in the supply-side.

\section{Problem Formalization}

Our problem is set in the top-$k$ recommendation during re-ranking stage, which considers a platform consisting of a group of users $\mathcal{U}=\{\bm{u}_i\in\mathbb{R}^d\}_{i=1}^m$ and a set of items $\mathcal{X}=\{\bm{x}_j\in\mathbb{R}^d,||\bm{x}||_2=1\}_{j=1}^n$. Table~\ref{Symbols} provides a list of symbols used in our paper. Items are described by feature vectors $\bm{x}_{j}\in\mathbb{R}^{d}$ and $d$ represents the number of semantic features. The system makes recommendations in the candidate list $\mathcal{C}_i \subseteq \mathcal{X}$ of individual $i$ by the ranking operator $\pi$ :
\begin{equation}\pi_i = \operatorname{rank}(\sigma_{i}(\bm{x}_{1}),\sigma_{i}(\bm{x}_{2}),...,\sigma_{i}(\bm{x}_{c})),\forall i\in[m].\end{equation}
It ranks items for each user $i$ using a personalized score function  $\sigma_{i}(\bm{x}) = \sigma(\bm{x}; \bm{u}_{i})= \bm{u}_{i}^\top \bm{x}$, which is capable of predicting the preference of a user over candidate items with the amount of $c$. Especially, the linear scoring function relies on user representation vectors $\bm{u}_{i}\in\mathbb{R}^{d}$ for each user $i\in[m]$. Overall, the goal of the system is to learn good $\sigma$ from data for re-ranking, where $\bm{u}_{1},..., \bm{u}_{m}$ are representations learned from the joint objective which is towards our debias setting for every user.

\begin{table}
  \caption{Symbols used in this paper.}
  \label{Symbols}
  \scalebox{1}{
  \begin{tabular}{l|l}
    \toprule
    Symbol & Description\\
    \midrule
    $\mathcal{X}$ & Semantic features of item set\\
    $\mathcal{U}$ & Representation of user set\\
    $\mathcal{U^*}$ & Ground-truth of user preference \\
    $\mathcal{C}$ & Users' candidate item list of recommendation\\
    $\sigma$ & Scoring function for ranking\\
    $\pi$ & Ranking operator towards recommendation \\
    $\mathcal{S}$ & Content-based  relevance simulator\\
    $r$ & Ground-truth of relevance generated by $\mathcal{S}$\\
    $f$ & Execution rules for strategic agent\\
    \midrule
    $d$ & Number of features of an item\\
    $k$ &  Number of items in the recommendation list\\
    $c$ &  Number of items in the candidate list\\
    $T$ & Number of retraining rounds in dynamics\\
    $\bm{P}$ &  Permutation matrix for ranking \\
    $\widehat{\bm{P}}$ & Relaxed permutation matrix\\
    $\tau$ & Temperature parameter for $\widehat{\bm{P}}$\\
    $\alpha$ & Scaling parameter of modification cost \\
    $\lambda$ & Regularization parameter for fair exposure \\

  \bottomrule
  \end{tabular}}
\end{table}

\section{Methodology}

In this section, we first detail key concepts and metrics in Section~\ref{Preliminaries}. This is followed by a discussion on how solutions have been tailored to address the non-differentiable nature of optimization metrics in Section~\ref{Differentiable Ranking}. Next, Section \ref{Strategic Agent} introduces the execution rule of the strategic agent. Finally, we formalize the process of dynamic learning and optimization in Section \ref{Dynamic Learning and Optimization}.

\begin{figure*}
  \centering
  \includegraphics[width=\linewidth]{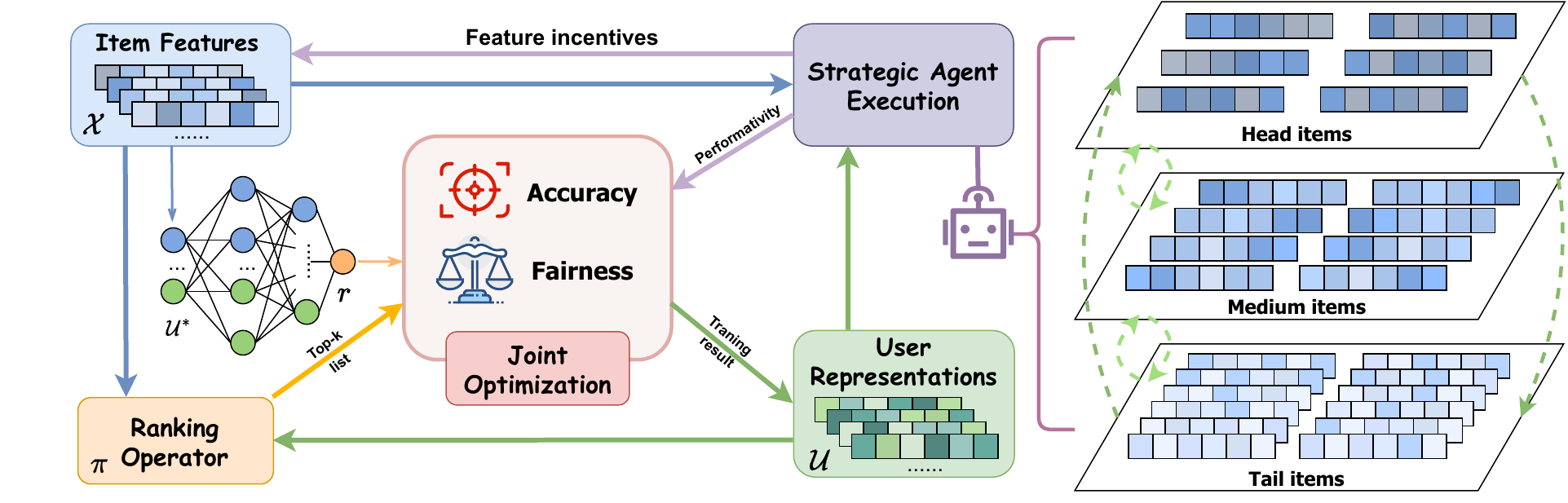}
  \caption{The framework of our re-ranking approach based on the strategic agent.}
  \label{framework}
\end{figure*}

\subsection{Preliminaries}
\label{Preliminaries}
In this section, we delineate the critical metrics for optimization and provide an overview of the dynamic training framework.

$\bullet$ \emph{\textbf{User Utility}.} In the context of evaluating ranking performance for a recommendation list, we define a list of items with their corresponding relevance scores for user $i$ as $\bm{r}_i=[r_1,r_2...,r_c]$. 
To represent different utilities for users over these items, we use ranking operator $\bm{\pi}=(\pi_1,\pi_2,...\pi_m)$, where $\pi(\ell)$ specifies the index in $r$ of the item that is ranked by $\bm{\pi}$ at position $\ell$. Then, the Discounted Cumulative Gain (DCG) at $k$-th position ($k \le c$) is defined as:
\begin{equation}\mathrm{DCG}_{@k}(\boldsymbol{r}_i,\pi_i)=\sum_{\ell=1}^k\frac{2^{r_{\boldsymbol{\pi}(\ell)}}-1}{\log(1+\ell)}.\end{equation}

Here, the numerator of terms captures the ``gain'' from the relevance of the item at position $\ell$. Conversely, the denominator serves to ``discount'' the value of this gain based on the rank $\ell$ reflecting the principle that items ranked lower contribute less to the cumulative gain due to the logarithmic discounting.
Then, the definition of the utility of users' recommendation list is as Normalised Discounted Cumulative Gain (NDCG):
\begin{equation}\mathrm{NDCG}_{@k}(\boldsymbol{r}_i,\pi_i)=\frac{1}{\mathrm{max} \mathrm{DCG}_{@k}(\boldsymbol{r}_i,\pi_i)}\mathrm{DCG}_{@k}(\boldsymbol{r}_i,\pi_i),\end{equation}
where $max\mathrm{DCG}_{@k}$ is the maximum possible value of $\mathrm{DCG}_{@\boldsymbol{k}}$, computed by the decreasing ground-truth relevance.

$\bullet$ \emph{\textbf{Item Exposure}.} We measure the inequality of item exposure in the recommendation list by the degree of imbalance between different relevance of items evaluated on the top-$k$ items. Specifically, we refer to the definition of Gini coefficient~\cite{do2022optimizing,wagstaff1991measurement,gastwirth1972estimation} in economics to measure the differences in exposure levels of different items. Formally, let top-$k$ relevance $\bm{r}_\pi$ (ranked by $\pi_i$) sorted in an ascending sequence, i.e., $r^1_{\pi}\leq r^2_{\pi}\leq...\leq r^k_{\pi}$. Then,
\begin{equation}
\mathrm{Gini}_{@k}(\boldsymbol{r}_i,\pi_i)=\frac{\sum_{\ell=1}^k(2\ell-k-1)r^\ell_{\pi}}{k\sum_{\ell=1}^kr^\ell_{\pi}}.
\label{Gini}
\end{equation}

The range of the Gini coefficient is between $0$ and $1$. The smaller $Gini_{@\boldsymbol{k}}(\boldsymbol{r}_i,\pi_i)$, the smaller the difference in relevance scores in the recommendation list, and vice versa. From the results of recommendations, there is a significant difference in the relevance score between tail items and head items. Our goal is to increase the exposure of tail products as much as possible while ensuring the performance of user recommendations. In other words, the Gini coefficient of the relevance score in the recommendation list should be relatively high, which can ensure that tail items have more fair exposure among user candidates.

$\bullet$ \emph{\textbf{Dynamic Environment}.}  Our study examines how learning with optimization influences user utility and item exposure across iterative rounds of retraining. Each iteration, denoted as the $t$-th round, is characterized by data  $\mathcal{D}^t=(\mathcal{U}^t,\mathcal{X}^t,\mathcal{C},\mathcal{U^*})$. During training, the simulator $\mathcal{S}(\mathcal{X}^t;\mathcal{U^*})$ is employed to generate the relevance $r$ between user preferences and item features, serving as a supervisory signal. Post each training iteration, content creators implement strategic modification for item features $\mathcal{X}^{t+1} = f(\mathcal{X}^t;\mathcal{U}^t)$ based on the trained representations of users. Then, the subsequent training iteration utilizes newly collected data and gets  $\mathcal{D}^{t+1}=(\mathcal{U}^{t+1},\mathcal{X}^{t+1},\mathcal{C},\mathcal{U^*})$. The dynamic framework is shown as Figure~ \ref{framework}. More details are provided in the following sections.

\subsection{Differentiable Ranking}
\label{Differentiable Ranking}
In this section, we propose differentiable ranking operators for measuring the user utility and the item exposure in the end-to-end learning framework for optimization. 

Given a relevance list between a user and their candidate items, $\bm{r}=[r_1,r_2,...,r_c]^\top$, our goal is to rank the items in the entire candidate set and calculate the NDCG and Gini values of the top-$k$ recommended item lists. In the re-ranking stage, the ranking operator $\pi_i$ used to generate the final list can be depicted as a permutation matrix, which makes ranking concerning NDCG and Gini coefficients non-differentiable. However, we can overcome this limitation by using the continuous relaxation method of permutation matrices~\cite{grover2019stochastic}, using differentiable and continuous operators to approximate deterministic ranking operations. First, the deterministic permutation matrix $\bm{p}_{\boldsymbol{r}}\in\mathbb{R}^{c \times c}$ defined by \cite{grover2019stochastic} is:
\begin{equation}\bm{P}_{\boldsymbol{r}}[p,q]=\begin{cases}1&if\ q=\arg\max[(c+1-2p)\boldsymbol{r}-\bm{A}_{\boldsymbol{i}}\mathbbm{1}]\\0& 
 otherwise\end{cases},\end{equation}
where $p$ and $q$ represent the rows and columns of the permutation matrix, $\mathbbm{1}$ denotes the column vector of all ones and $\bm{A}_{\boldsymbol{i}}[p,q] = |r_p-r_q|$ which is the matrix
of absolute pairwise differences of the elements in the list. Then, the $\arg\max$ operator can be replaced by softmax\cite{jang2016categorical} to obtain a
continuous relaxation:
\begin{equation}\widehat{\bm{P}}_{\boldsymbol{r}}[p,:](\tau)=\text{softmax}[((c+1-2p)\boldsymbol{r}-\bm{A}_{\boldsymbol{i}}\mathbbm{1})/\tau],\end{equation}
where $\tau > 0$ is a temperature parameter, serving as the level of smoothness in the approximation. A lower value of 
$\tau$ nudges the relaxed permutation matrix towards a more deterministic nature. This relaxation is continuous and differentiable in relation to the elements of $\bm{r}$, ensuring a balance between computational tractability and the fidelity of the approximation.

Following \cite{pobrotyn2021neuralndcg}, we implemented Sinkhorn scaling~\cite{sinkhorn1964relationship} to refine the permutation matrix to get $scale(\widehat{\bm{P}}_{\boldsymbol{r}})$, ensuring that both rows and columns become probabilistically distributed with a total of one. This approach effectively diminishes the gradient variance around the discontinuity points of the optimization metrics, thereby facilitating the efficiency of the gradient descent method in optimization.

Utilizing the ranking scores derived from the  $\pi$ along with the ground-truth relevance $\bm{r}$, we employ $scale(\widehat{\bm{P}}_{\boldsymbol{r}})$ to ascertain the approximate values of the NDCG and the Gini coefficient for the current recommendation list. This process facilitates the creation of a differentiable ranking operator. In terms of user utility, we adopt the approach as delineated by \cite{pobrotyn2021neuralndcg}, which yields:
\begin{equation}
\label{DR-NDCG}
\begin{aligned}
&\operatorname{DR-NDCG}_{@k}(\boldsymbol{r},\pi)(\tau_1) \\
&=(\sum_{\ell=1}^k g(r_{(\ell)})\cdot d(\ell))^{-1}\cdot \sum_{\ell=1}^k[\operatorname{scale}(\widehat{\boldsymbol{P}})\cdot g(\boldsymbol{r})]_\ell\cdot d(\ell), \\
\end{aligned}
\end{equation}
where $g(x)=2^{x}-1$ and $d(x)=\frac1{\log_2(x+1)}$. $\ell$ is the index of the ranking position. The notation $r_{(\ell)}$ represents the relevance score of the $\ell$-th item after sorting $\bm{r}$ in descending order.

In terms of item exposure, we have developed a new differentiable ranking operator for inequality measure. That makes it feasible for gradient-based optimization in the following section. Formally, let $\bm{r}_\pi$ denote the ground-truth relevance of items the in recommendation list. Then:
\begin{equation}
\label{DR-Gini}
\begin{aligned}
&\operatorname{DR-Gini}_{@k}(\boldsymbol{r},\pi)(\tau_2) \\
&=\frac{1}{\bar{\bm{r}}_\pi k^2}\sum_{p=1}^k\sum_{q=1}^k|[\operatorname{scale}(\widehat{\boldsymbol{P}})\cdot \boldsymbol{r}]_p-[\operatorname{scale}(\widehat{\boldsymbol{P}})\cdot \boldsymbol{r}]_q|, \\
\end{aligned}
\end{equation}
where $\bar{\bm{r}}_\pi = \frac{1}{k}\sum_{i=1}^k r_{\pi}^{i}$. To facilitate the formula inspired by pairwise ranking and analysis~\cite{cao2007learning,cheng2019alpha,wu2023exploring, wu2023learning}, we use variant formulas based on Eq.(\ref{Gini}) to describe the difference between pairwise elements. This can also better reflect the connotation of the Gini coefficient index we proposed. It can measure the unfairness of exposure caused by relevance discrepancies in the candidate list.

\subsection{Strategic Agent}
\label{Strategic Agent}
Considering the distinct popularity levels of items, the number of audiences each item possesses also differs. Initially, we define the utility function of an item $s(x_j)$ based on the average score of its rankings within the target demographic for item $x_j$. We denote $\mathcal{U}_j$ as the set of users whose candidate list contains item $x_j$. Then,
\begin{equation}
s(x_j) = \hat{w}_j^\top x_j = \frac{1}{|\mathcal{U}_j|} \sum_{u \in \mathcal{U}_j} \sigma(x_j; u). 
\end{equation}

In order to compare the different sum quantity of user representations within the same range, a norm constraint is necessary for them~\cite{eilat2023performative,hron2022modeling}. Therefore, we normalize the $l_2$ norm of the summation of user representations specific to item $x_j$, denoted as
$$\hat{w}_j=\frac{w_j}{||w_j||_2}, \operatorname{where} w_j=\frac{1}{|\mathcal{U}_j|}\sum_{u\in \mathcal{U}_j}u .$$

Then, we define the execution rules of the strategy agent in systems. In alignment with the strategic classification framework established by \cite{hardt2016strategic}, we posit that content creators have the ability to modify their features of items, though this incurs a cost, as a reaction to the established predictive model. Considering a predefined cost function $cost(x_j,x^{\prime})$, these creators adjust their items following an optimal response strategy: 
\begin{equation}
\Delta_f(x_j) = \operatorname*{argmax}_{x^{\prime}:\|x^{\prime}\|_2=1} \left[ s(x^{\prime}) - \alpha \cdot cost(x_j, x^{\prime}) \right],
\label{reponse}
\end{equation}
where $cost(x,x^{\prime}) = \|x-x^{\prime}\|_2^2$ denotes the cost of modification. It measures the degree of deviation after execution, whose scaling parameter is $\alpha \ge 0$. The smaller $\alpha$ illustrates that the strategic agent employs the more aggressive policy for modification. This results in item features deviating more significantly from their original semantic information and vice versa.
The implementation of normalization ensures that modified items maintain their unit l2 norm, preserving their dimensional integrity.

In Eq.(\ref{reponse}), the argmax function is non-differentiable, and the term $\Delta_f$ relies on the optimized representations concerning both accuracy and fairness objectives. That presents challenges for the end-to-end implementation of the strategic agent.

To capitalize on the performativity of prediction, we have derived a differentiable formulation in terms of $\Delta_f$. This formulation elucidates the viability of execution by the strategic agent within a gradient optimization framework. Initially, we expand Eq.(\ref{reponse}) using a Lagrangian expression Eq.(\ref{Lagrangian}), where $\gamma$ represents the Lagrange multiplier. The equations are as below:
\begin{equation}L(x^{\prime},\gamma) = w^Tx^{\prime}-\alpha\|x^{\prime}-x\|_2^2+\gamma(\|x^{\prime}\|_2^2-1)\label{Lagrangian}.\end{equation}

Subsequently, by applying the Karush-Kuhn-Tucker (KKT)\cite{wu2007karush} conditions, we arrive at the final formulation:
\begin{equation}
 \Delta_f(x_j) =f(x_j;\hat{w}_j) = \frac{\hat{w}_j+2\alpha x_j}{\|\hat{w}_j+2\alpha x_j\|_2}.
\label{KKT}
\end{equation}

A detailed proof is furnished in Appendix \ref{PROOF}, where the application of differentiable execution facilitates the derivation of a readily solvable form, thereby enhancing the efficacy of dynamic optimization in the subsequent section.

\subsection{Dynamic Learning and Optimization}
\label{Dynamic Learning and Optimization}
\subsubsection{Muti-round Dual-objective Optimization.} 
The re-ranking stage determines which items appear in the top-$k$ recommendation list finally, directly influencing the exposure of items from the producer side. It is worth noting that fully pursuing the fairness of exposure will worsen the original accuracy of the recommendation. To this end, we apply a dual-objective optimization to the post-processing re-ranking algorithm in every iteration of dynamic interaction.

This approach ensures the user's top-$k$ recommendation list not only remains effective but also promotes the visibility of long-tail products.
The differentiable joint optimization objective in the candidate list is defined as:

\begin{equation}\max_{u_i\in\mathcal{U}}\frac1m\sum_{i=1}^m[\operatorname{NDCG}_{@k}(\boldsymbol{r}_i,\pi_i)+\lambda\operatorname{Gini}_{@k}(\boldsymbol{r}_i,\pi_i)]\label{original formula},\end{equation}
where $\bm{r}_i = \mathcal{S}(\mathcal{C}_i;\bm{u}_i^*)$ is ground-truth relevance score quantifying the degree of relatedness between user $i$ and items in his candidate list.   $\pi_i=\operatorname{rank}(\sigma(\mathcal{C}_i;\bm{u}_i))$ is a ranking operator to select items for user $i$. The regularization parameter $\lambda$ is the controlling degree for fair exposure of items. Essentially, We denote $\bm{u}_i^*$ as users' ground-truth preference which consists of total candidate information:
$$\bm{u}_i^*=\frac{\bm{u}_i^\prime}{\|\bm{u}_i^\prime\|_2}, \operatorname{where} \bm{u}_i^\prime=\frac1{|\mathcal{C}_i|}\sum_{x\in \mathcal{C}_i}x .$$

Constrained with unit $\ell2$ norm, $\bm{u}_i^*$ can be used as a supervisory signal in every round of the optimization process. After the current $t$-th round ends, we get trained representation $\mathcal{U}^t$. The strategy agent undertakes the modification of item features: $\mathcal{X}^{t+1}=f(\mathcal{X}^t;\mathcal{U}^t)$, i.e., $x_j^{t+1}=\Delta_f(x_j^{t})$. Then, the relevance score is re-calculated by content-based simulator $\mathcal{S}$ denoted as $\bm{r}_{i}^{t+1}=\mathcal{S}(\mathcal{C}_{i}^{t+1};\bm{u}_i^*)$, where $\mathcal{C}_{i}^{t+1}\subseteq \mathcal{X}^{t+1}$, so that we can use $\bm{r}_{i}^{t+1}$ to re-train user representations at $t$+1-th round.

\begin{algorithm}[t]
\caption{Multi-round Training Process}
\begin{algorithmic}[1] 
\State Training procedure at time period $t$;
\State \textbf{Input:} Item features $\mathcal{X}^t$, initialized user representations $\mathcal{U}$, the relevance simulator $\mathcal{S}$, hyperparameters $\lambda, \alpha, \tau$;
\State \textbf{Output:} Updated user representations $\mathcal{U}^t$, modified item features $\mathcal{X}^{t+1}$;
\For{epoch}
\State Get $\Delta_{f}(\mathcal{X}^t)=f(\mathcal{X}^t;\mathcal{U}^t)$  based on Eq.(\ref{KKT});
    \For{user $i\in{\{1,2,...,m\}}$}
    \State Get $\mathcal{C}_i^t \subseteq \mathcal{X}^t$ and $\Delta_{f}(\mathcal{C}_i^t) \subseteq \Delta_{f}(\mathcal{X}^t)$
    \State Use $\widehat{\bm{P}}$ based on $\pi_i$, $\Delta_{f}(\pi_i)$ to generate the top-$k$ list;
    \State Use $\mathcal{S}$ to get $\bm{r}_i$, $\Delta_{f}(\bm{r}_i)$ ;
    \State Calculate $\operatorname{NDCG}_{@k}(\boldsymbol{r}_i,\pi_i)$ as Eq.(\ref{DR-NDCG})
    \State Calculate anticipated $\operatorname{Gini}_{@k}(\Delta_{f}(\boldsymbol{r}_i),\Delta_{f}(\pi_i))$ as Eq.(\ref{DR-Gini})
    \EndFor
    \State Calculate total optimization loss as Eq.(\ref{loss});
    \State Update $\mathcal{U}^t$ via gradient-descent;
\EndFor
\State Get $\mathcal{X}^{t+1}=f(\mathcal{X}^t;\mathcal{U}^t)$ based on Eq.(\ref{KKT})
\State \textbf{return} $\mathcal{U}^t$, $\mathcal{X}^{t+1}$
\end{algorithmic}
\label{algorithm process}
\end{algorithm}

\subsubsection{Agent-based Strategic Learning.} 
The system gains potential power to shape incentives for fairness based on content creators' reliance on the learned function $\sigma$ for utility. That means the update of strategic agents has the potential to promote fair exposure.

The original optimization process shown as Eq.(\ref{original formula}) ensures fairness in the current learning iteration. That is reactive, specifically adapted to conditions from the preceding time step. Recognizing that the execution function $f$ encourages content creators to alter items, we suggest a shift towards proactive promotion of fairness by foreseeing and adapting to their strategic response. Specifically, we utilize the performativity of prediction to carry out a forward-looking regularization of optimization for fair exposure:
\begin{equation}\max_{u_i\in\mathcal{U}}\frac1m\sum_{i=1}^m[\operatorname{NDCG}_{@k}(\boldsymbol{r}_i,\pi_i)+\lambda\operatorname{Gini}_{@k}(\Delta_{f}(\boldsymbol{r}_i),\Delta_{f}(\pi_i))]\label{agent-based}.\end{equation}

As for regularization term , we replace $\mathcal{C}_i$ with the anticipated form $\Delta_{f}(\mathcal{C}_{i})=\{\Delta_{f}(x)\}_{x\in \mathcal{C}_{i}}$. Then, $\bm{r}_i$ is also promptly replaced with $\Delta_{f}(\bm{r}_i)$ induced by $\mathcal{S}$: $\Delta_{f}(\bm{r}_i)=\mathcal{S}(\Delta_{f}(\mathcal{C}_{i});\bm{u}_i^*)$. Similarly, the ranking operator is also reshaped by $f$, since the change of $\mathcal{C}_i$, denoted as $\Delta_{f}(\pi_i)=\operatorname{rank}(\sigma(\Delta_{f}(\mathcal{C}_{i});\bm{u}_i))$. Thus, the response of strategy agents can be fed back to the current optimization process in real-time, thereby giving the item set more incentives for non-popularity characteristics.

Further, we get the intuitive form of optimization loss based on Eq.(\ref{agent-based}) to perform training of user representation $\bm{u}_i$ at $t$-th round. Specifically, the loss function for agent-based strategic learning is shown as follows:
\begin{equation}
\label{loss}
\begin{aligned}
\mathcal{L}(\mathcal{D}_t)=&-\frac{1}{m}\bigg[ \sum_{i=1}^m\operatorname{DR-NDCG}_{@k}(\boldsymbol{r}_i,\pi_i)(\tau_1)\\
&+\lambda\sum_{i=1}^m\operatorname{DR-Gini}_{@k}(\Delta_{f}(\boldsymbol{r}_i),\Delta_{f}(\pi_i))(\tau_2)\bigg].
\end{aligned}
\end{equation}

Eq.(\ref{loss}) optimizes for the current user utilities and anticipated item exposure, which can fully leverage the role of strategic agents in the dynamic learning process. For the complete algorithm framework of muti-round training, see Algorithm \ref{algorithm process}.

\section{Experiments}
In this section, we undertake a series of experiments utilizing two real-world datasets to assess the efficacy of our suggested re-ranking approach. Our objective of the experiment is to address the ensuing research questions (RQs):
\begin{itemize}[leftmargin=*]
\item \textbf{RQ1:} How effective is the strategic agent in balancing user utility and fair exposure of items in a dynamic setting?
\item  \textbf{RQ2:} How do different hyper-parameter settings (e.g. $\lambda$, $\alpha$) during optimization affect the re-ranking performance?
\item  \textbf{RQ3:} How does our re-ranking approach eliminate the popularity bias over the long term?
\end{itemize} 

\subsection{Content-based Simulator}
Before the main experiments, we train two simulators to get users' ground truth relevance of items with varying features over time. Then, we can implement multiple rounds of training.

\subsubsection{Datasets. }
We utilized a publicly available dataset and a real-world industrial dataset for training simulators, which are used for the following main experiment. Table \ref{statistics} shows the necessary statistics for these two datasets.

\textbf{The Yelp restaurant dataset.} Yelp provides information on a wide variety of businesses and our study specifically concentrates on the restaurant sector. We train $\mathcal{S}$ on this generated by users having at least 50 restaurant reviews following \cite{eilat2023performative}. These amount to 1,377 users, 22,197 restaurants and 113,852 reviews. The number of restaurant informative features is \textbf{43}. 

\textbf{The industrial dataset.} As for the industrial, we selected out users who specialize in targeting laptops' consumption and have at least 10 clicks on different items based on their behavior logs in the online shopping mall. The data time range is from January 2023 to December 2023. It has 11238 users, 1214 items, and 270259 clicks. The number of laptop feature dimensions is \textbf{24}. We will provide specific information about the features in Appendix \ref{feature description}.

\begin{table}[h]
\centering
\caption{Statistics of the Yelp and Industrial dataset.}
\label{statistics}
\begin{tabular}{@{\extracolsep{7pt}}lcc@{\extracolsep{7pt}}}
\toprule
Data set     & Yelp restaurant        & Industrial   \\
\midrule
\#Users        & 1,377 (242)  & 11,238 (351) \\
\#Items        & 22,197 (1,103) & 1,214  (951) \\
\#Interactions & 113,852      & 270,259      \\
Density     & 0.00372      & 0.0198      \\
\bottomrule 
\end{tabular}
\end{table}

\subsubsection{Design and Training. }
We design the architecture of the simulator, which can capture non-linear relations between user and item features found in the data. Specifically, we configure the simulator as an MLP featuring ReLU activations. Our architectural choice includes five layers. The initial layer is designed to accept 2$d$ inputs—comprising $d$ item features and $d$ user features—and produces 4$d$ outputs. Subsequently, the output dimension of each successive layer is halved, ensuring a systematic reduction in dimensionality throughout the network. 

For each user $i$, we derive the ground-truth user features 
by aggregating the features of all items reviewed (clicked) by user $i$. Then, for Yelp, take the restaurants reviewed as positive and those that have not been reviewed but closest to the geographical locations that have been reviewed as negative samples. For industrial, Randomly select 10 items from the user's click list as positive samples, and 10 items outside the click list as negative samples. The final $\mathcal{S}$ trained by BCE loss (70-20-10) for two datasets achieves 71.8\% and 90.7\% accuracy respectively on the held-out test set.

\subsection{Experimental Settings}
\subsubsection{Implementation Details.}
\begin{figure}[htbp]
    \centering
    \begin{subfigure}[b]{0.48\linewidth}
        \includegraphics[width=\linewidth]{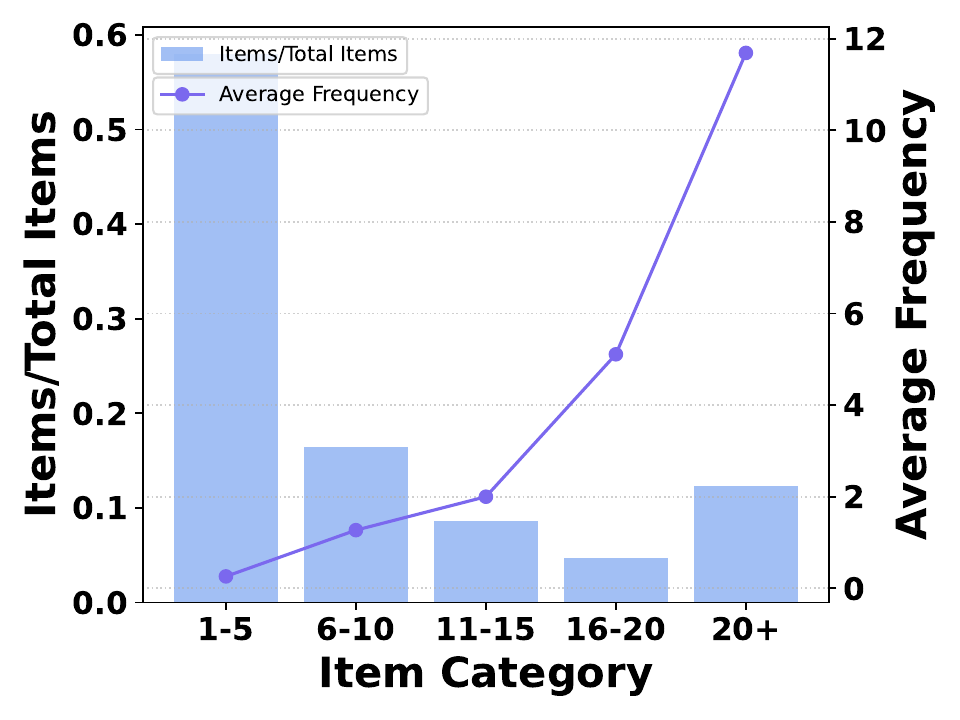}
        \caption{Yelp dataset}
    \end{subfigure}
    \hfill 
    \begin{subfigure}[b]{0.48\linewidth}
        \includegraphics[width=\linewidth]{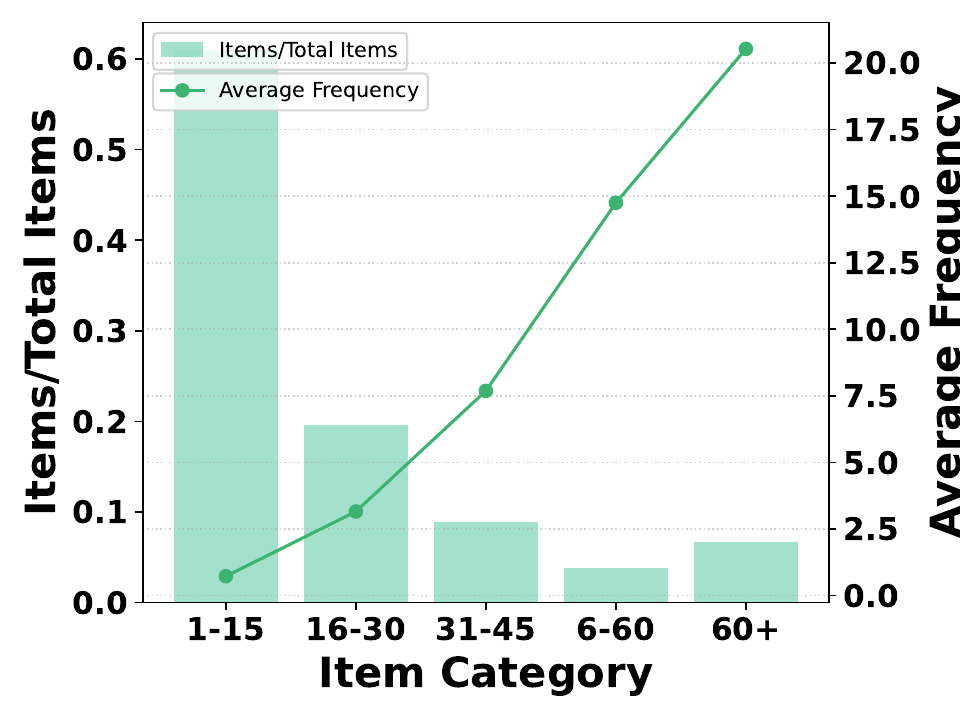}
        \caption{Industrial dataset}
    \end{subfigure}
    \caption{Average frequency of different item categories recommended by Content-based Simulator when $k=10$.}
    \label{bias_illustration}
\end{figure}

In the main experiment, we focus on active users in platforms. Specifically, we select data from a city (Vancouver) in Yelp to explore the effect of popularity bias better. Initially, we filter out users with more than c interactions and select $c$ items to form each user's candidate list. Subsequently, items are classified into five categories based on their frequency of appearance across all candidate lists of users. For Yelp, setting c=40 results in 242 active users and 1103 items, with the item categories being [`1-5', `6-10', `11-15', `16-20', `20+'] and their respective count list is [634, 180, 94, 51, 135]. For industrial, setting c=50 results in 351 active users and 951 items, with the item categories being [`1-15', `16-30', `31-45', `46-60', `60+'] and their respective count list is [580, 186, 85, 36, 64]. Notably, original top-$k$ lists reveal that items with higher popularity are more frequently recommended, as shown in Figure \ref{bias_illustration}, which indicates a tendency for $\mathcal{S}$ to assign higher relevance to items of greater popularity. In particular, the average Gini coefficients across users' original top-$k$ lists are 0.0149 (Yelp) and 0.0079 (industrial), respectively.

We set $k$=10 and T=10, randomly selecting $c$-10 items from $\mathcal{C}_i$ of per user for training in each round, and test using total $c$ items. The temperature coefficients were set to $\tau_1$=0.1 and $\tau_2$=1. We conducted 100 epochs per round, with a learning rate of 0.1, a batch size of 64 on Yelp, and a batch size of 32 on industrial. All codes are programmed in Python 3.9.17 and PyTorch 2.0.1.

\begin{figure*}[htbp] 
\captionsetup{belowskip=0pt, aboveskip=4pt} 
    \centering
    \includegraphics[width=\textwidth,trim=0 50 0 20, clip]{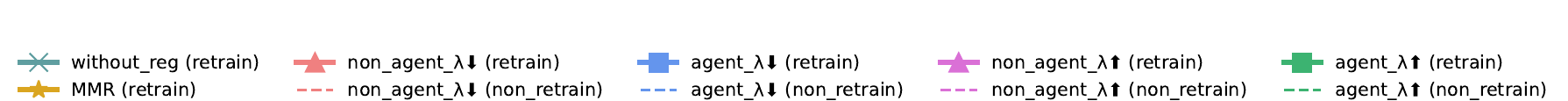} 
    \label{legend}
    \begin{subfigure}[b]{0.32\textwidth} 
        \includegraphics[width=\textwidth]{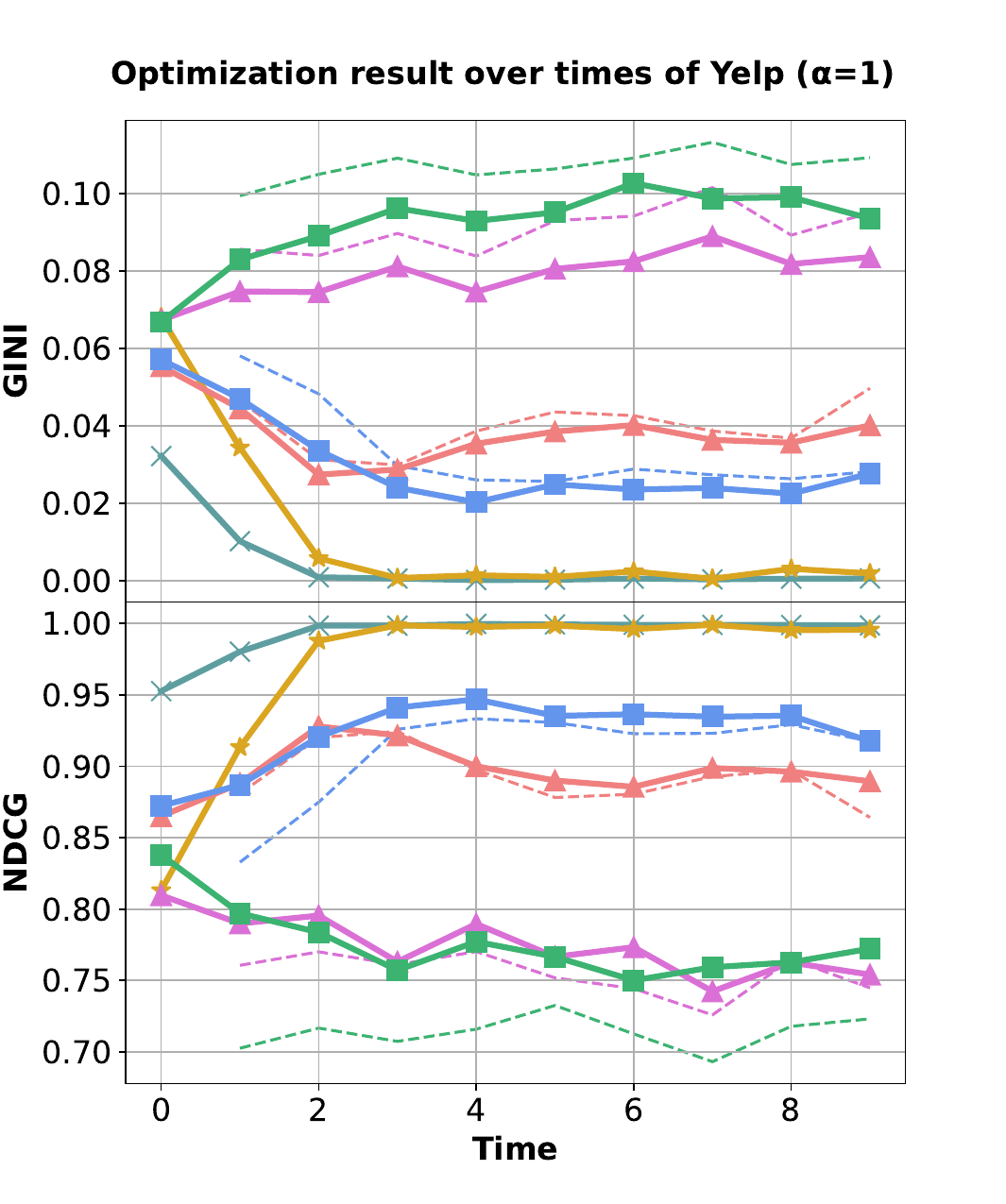} 
        \caption{$\alpha = 1$}
        \label{fig:Yelp1}
    \end{subfigure}
    \hspace{-5pt} 
    \begin{subfigure}[b]{0.32\textwidth}
        \includegraphics[width=\textwidth]{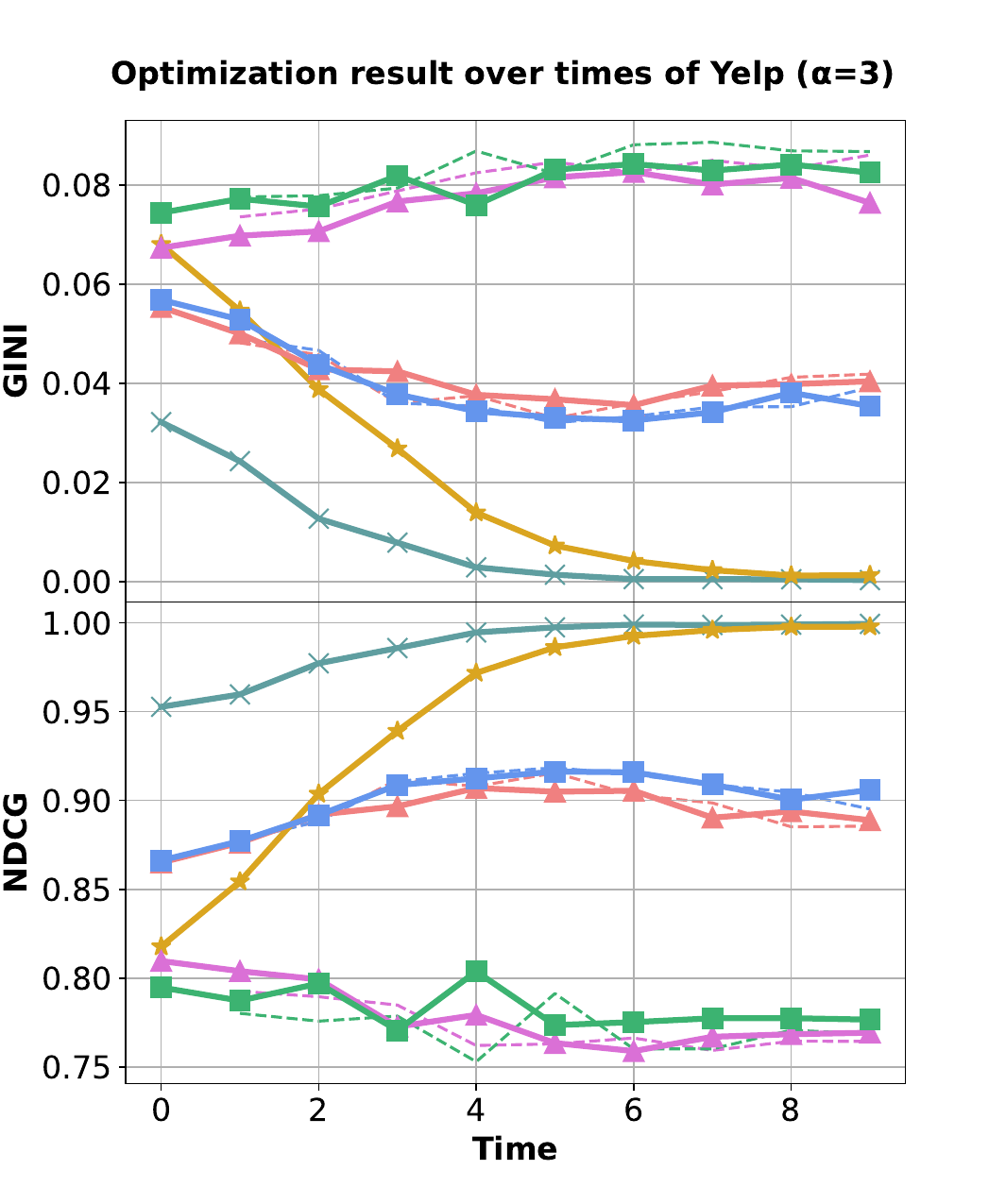}
        \caption{$\alpha = 3$}
        \label{fig:Yelp2}
    \end{subfigure}
    \hspace{-5pt} 
    \begin{subfigure}[b]{0.32\textwidth}
        \includegraphics[width=\textwidth]{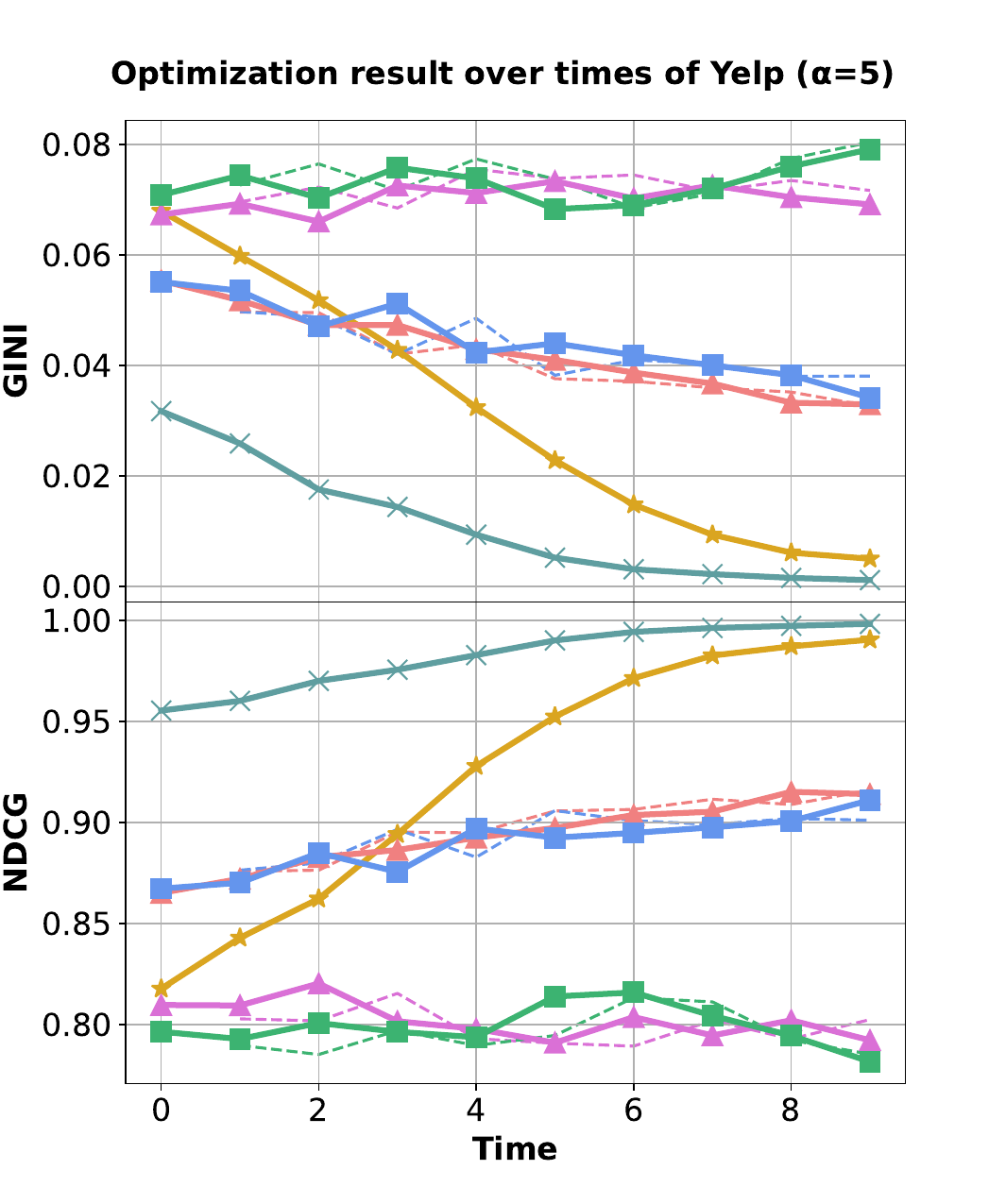}
        \caption{$\alpha = 5$}
        \label{fig:Yelp3}
    \end{subfigure}
    \caption{Optimization results over times of Yelp datasets with different $\alpha$.}
    \label{fig:Yelp_result}
\end{figure*}

\begin{figure*}[htbp] 
\captionsetup{belowskip=0pt, aboveskip=4pt} 
    \centering
    \begin{subfigure}[b]{0.32\textwidth} 
        \includegraphics[width=\textwidth]{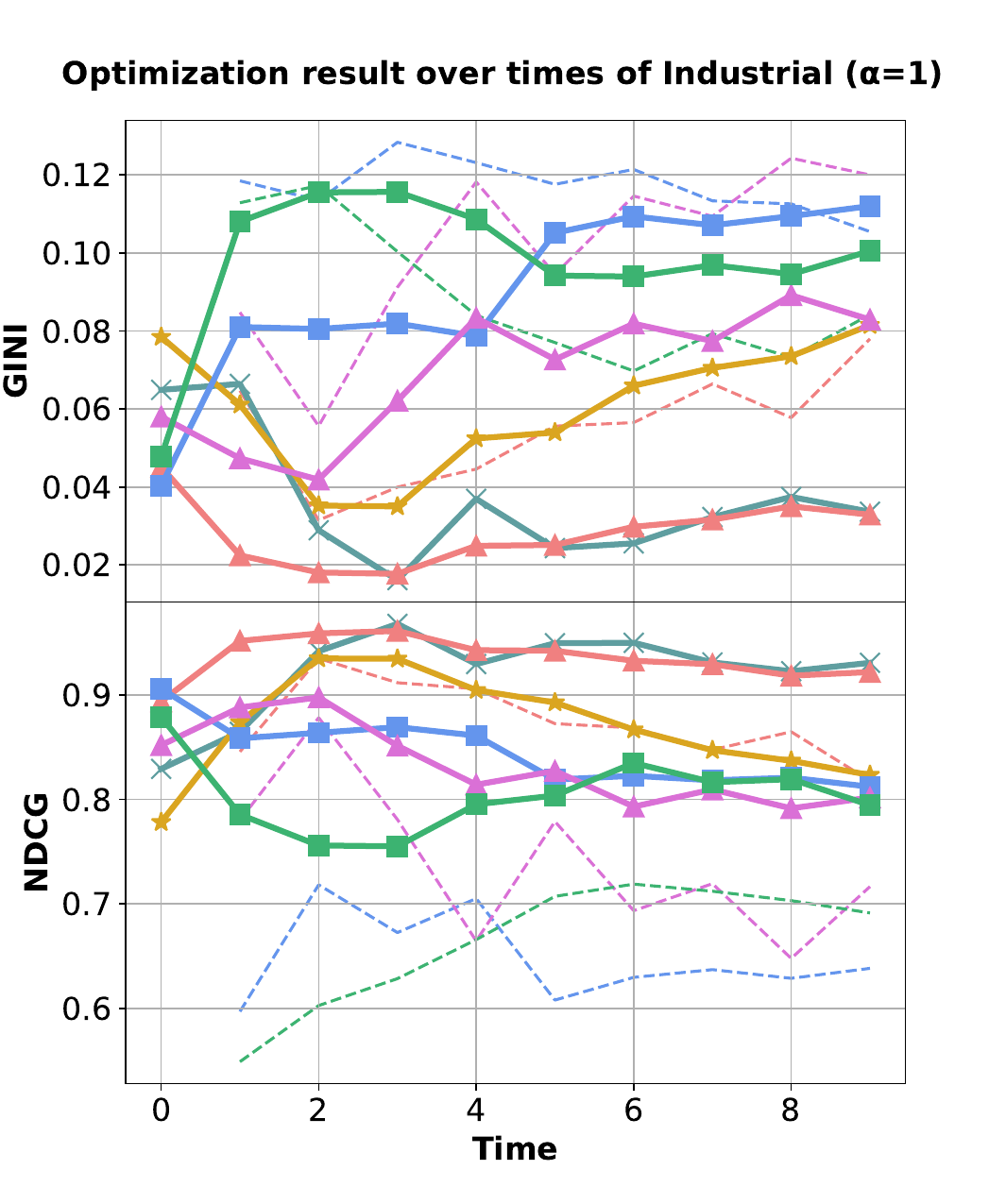} 
        \caption{$\alpha = 1$}
        \label{fig:Industrial1}
    \end{subfigure}
    \hspace{-5pt} 
    \begin{subfigure}[b]{0.32\textwidth}
        \includegraphics[width=\textwidth]{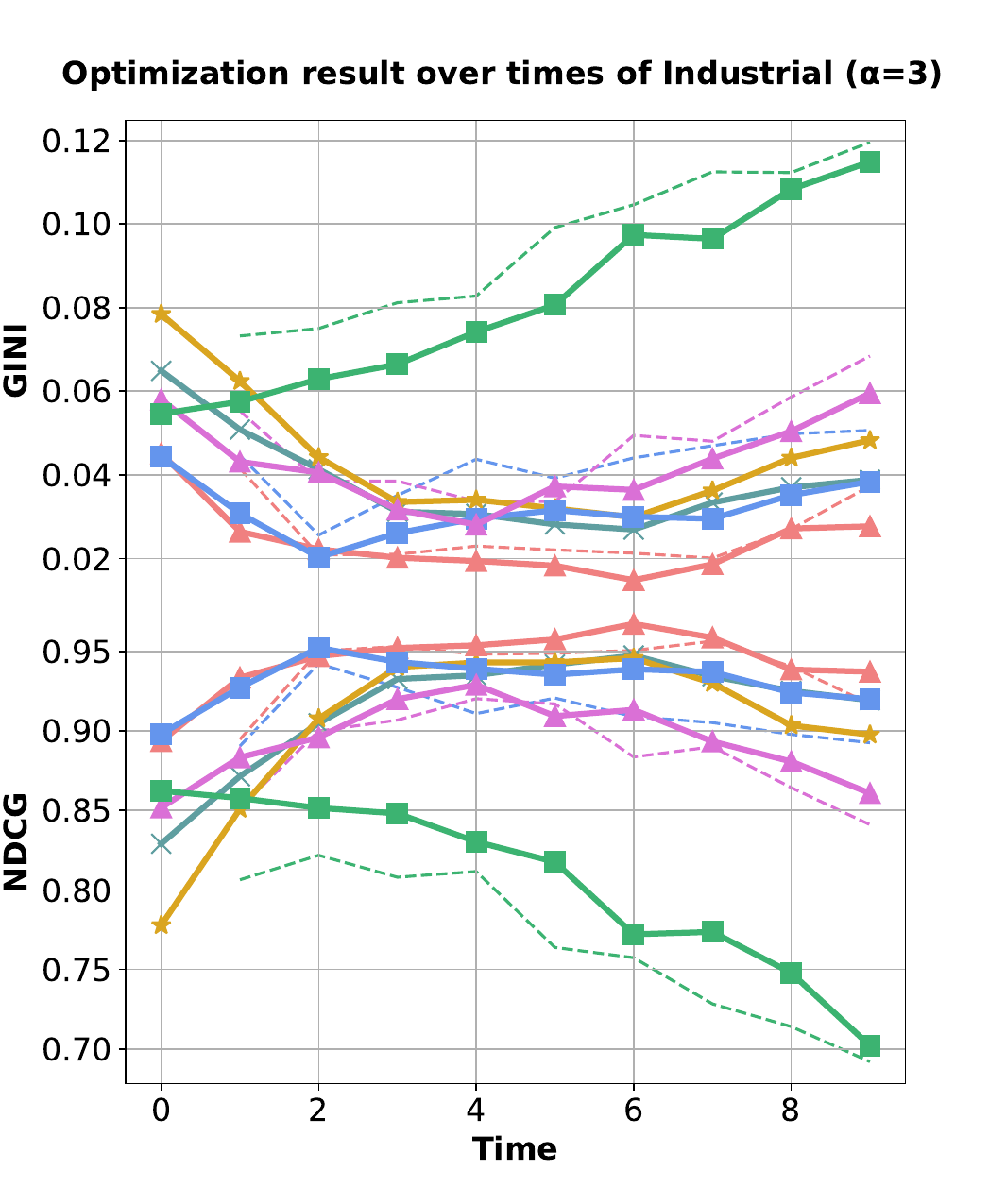}
        \caption{$\alpha = 3$}
        \label{fig:Industrial2}
    \end{subfigure}
    \hspace{-5pt} 
    \begin{subfigure}[b]{0.32\textwidth}
        \includegraphics[width=\textwidth]{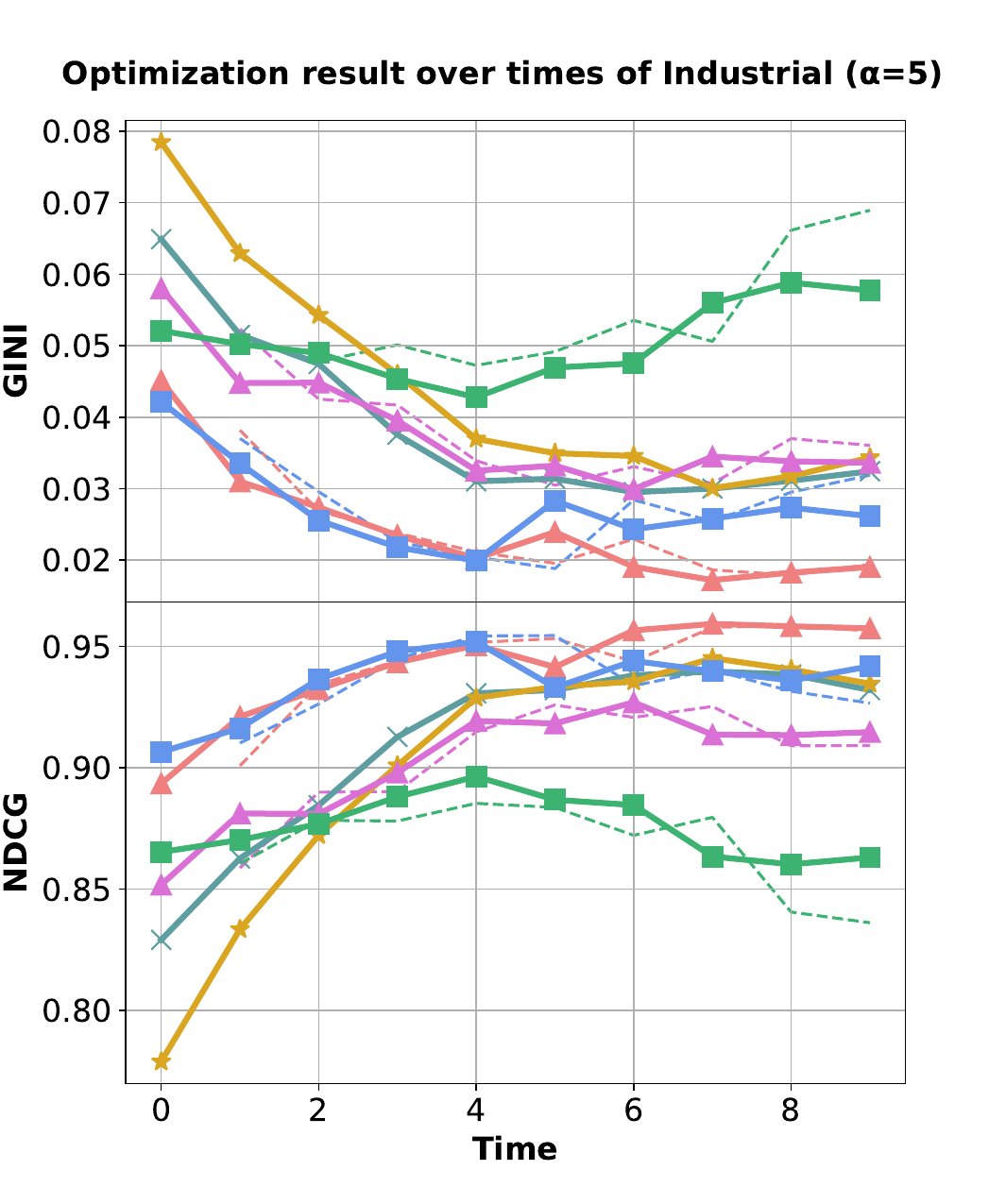}
        \caption{$\alpha = 5$}
        \label{fig:Industrial3}
    \end{subfigure}
    \caption{Optimization results over times of industrial datasets with different $\alpha$.}
    \label{fig:Industrial_result}
\end{figure*}

\subsubsection{Compared Methods} To compare the performance of different re-ranking methods in dynamic environments, we set up multiple baselines to verify the effectiveness of our proposed model:
\begin{itemize}[leftmargin=*]
\item The \textit{agent-based} optimization approach: it carries out a forward-looking regularization for fair-exposure of items as Eq.(\ref{agent-based});
\item The \textit{non-agent} optimization approach: it focuses on regularization at the current round without anticipated inputs as Eq.(\ref{original formula});
\item The \textit{accuracy-only} baseline: it pays no attention to the fairness of item exposure and each training round only seeks to maximize user utilities (by setting $\lambda=0$ as Eq.(\ref{loss}));
\item \textit{MMR}~\cite{carbonell1998use}: a traditional re-ranking approach using the Maximal Marginal Relevance procedure, which balances accuracy and diversity in top-$k$ lists based on the accuracy-only baseline; 
\item The \textit{non-retraining} approach: items are modified according to the model from the previous round and performance is tested on those changes without retraining the model. 
\end{itemize}

\subsection{Experimental Results (RQ1\&2)}
The outcomes of our experiments conducted across two datasets are depicted in Figure \ref{fig:Yelp_result} and Figure \ref{fig:Industrial_result}. we have employed different $\lambda$ to showcase our findings, ensuring that the ultimate NDCG remains above 0.7 while avoiding overfitting. Specifically, for Yelp, $\lambda$ is set 5 ~($\downarrow$) and 10 ($\uparrow$); for Industrial, $\lambda$ is set 2 ($\downarrow$) and 4 ($\uparrow$).

\subsubsection{The Effect of Strategic Agents (RQ1). }
\label{The Effect of Strategic Agents (RQ1).}
Firstly, we analyze the system's performance with and without fair-exposure. The results show that the system always makes the Gini coefficient higher after the fairness regularization than before. Furthermore, the traditional re-ranking method (MMR), when confined to the accuracy-only approach, fails to sustain fair exposure across increasing training iterations. This failure stems from the accuracy-only approach driving the system into homogenization, where item features in lists are overshadowed by those of more popular items, leading to a convergence of NDCG towards 1.

As for retraining and non-retraining approaches, it is evident that in every iteration, the NDCG values derived from the non-retraining approach always fall below those attained through retraining, depicted in Figure \ref{fig:Yelp1}. This disparity arises because retrained models more accurately encapsulate the information refined during the current iteration of optimization. For industrial, where each user has more candidates, the discrepancy between retraining and non-retraining is markedly pronounced especially when $\alpha$ is small. 

Then, our analysis extends to evaluating the influence of anticipatory regularization on fair-exposure optimization, contrasting agent-based ($\square$) and non-agent ($\triangle$) methods. In scenarios where $\lambda$ is relatively high, agent-based methods always surpass non-agent approaches in optimizing the Gini coefficient. This suggests that, under comparable conditions, optimization strategies that incorporate looking-forward intervention $(\Delta_{f}(\boldsymbol{r}_i),\Delta_{f}(\pi_i))$ (Eq.(\ref{agent-based})) more rapidly align the system towards equitable exposure. Remarkably, even at lower $\lambda$ values, agent-based methodologies demonstrate superior performance in initial training rounds. The comparison results illustrate that utilizing the performativity of prediction can realize fair-exposure optimization for debias objectives more effectively and efficiently.

Most importantly, contrary to intuitive expectations that increased fair exposure correlates with higher user utility deterioration—implied by a direct trade-off between Gini coefficient enhancement and NDCG reduction—the agent-based approach defies this phenomenon. It achieves superior Gini coefficients without compromising, and in fact, while enhancing NDCG values compared to non-agent benchmarks, as illustrated in Figure \ref{fig:Yelp1}, Figure~ \ref{fig:Yelp2} and Figure \ref{fig:Industrial1} (after the sixth time step) for $\lambda\ (\uparrow)$. This achievement suggests that the agent-based method adeptly identifies and leverages beneficial features of less popular items within a limited range of time steps, thereby balancing the enhancement of tail item visibility with the concurrent improvement of user utilities.

\subsubsection{Hyper-parameter Analysis (RQ2). } 
In this subsection, the role of regularization parameter $\lambda$ and scaling parameter of modification costs $\alpha$ is identified in our dynamic setting.

Focusing on $\lambda$, larger values significantly bolster the system's capability for fair-exposure optimization, manifesting in a preference to substitute popular features in the initial top-$k$ list with less popular attributes. This adjustment encourages strategic agents to increasingly incentivize tail item features within the recommendation list across successive training rounds. Consequently, the larger $\lambda$ results in a progressive elevation of the Gini coefficient and a corresponding decline in NDCG values as training advances. However, it is important to note that this decline is not absolute and is dependent on the value of $\alpha$, as explained in the latter half of the subsection \ref{The Effect of Strategic Agents (RQ1).}. Conversely, lower lambda values facilitate a reduction in the Gini coefficient and expedite the system's drift toward homogenization, especially noted in the industrial dataset.

As for $\alpha$, the higher value imply a greater cost for strategic agents to modify item features, leading to a reduction in feature incentives across disparate item categories. That enables larger $\alpha$ to moderate the system's iterative evolution. For illustration, contrasts drawn between Figure \ref{fig:Yelp2} and Figure \ref{fig:Yelp3}, as well as Figure \ref{fig:Industrial2} and Figure~ \ref{fig:Industrial3}, reveal that at lower values of $\lambda$ , the Gini coefficient diminishes progressively with each training round. However, at the $\alpha$ setting of 5, the Gini coefficient remains consistently higher than at the $\alpha$ setting of 3, thereby tempering the momentum towards homogenization. In contrast, with higher lambda values, although the Gini coefficient escalates with each round, $\alpha=5$ ensures a lower Gini coefficient compared to $\alpha=3$, effectively moderating the drive towards enhanced fair-exposure. Furthermore, the value of $\alpha$ can impact the disparity between retraining and non-retraining approaches. By comparing the three sub-figures in Figure \ref{fig:Yelp_result} and Figure \ref{fig:Industrial_result}, it can be intuitively seen that the larger the value of parameter $\alpha$, the smaller the difference between the two methods.

\subsection{Effectiveness Results of Debias (RQ3)}
\begin{figure}[htbp]
    \centering
    \begin{subfigure}[b]{0.48\linewidth}
    \captionsetup{belowskip=0pt, aboveskip=0pt}
        \includegraphics[width=\linewidth]{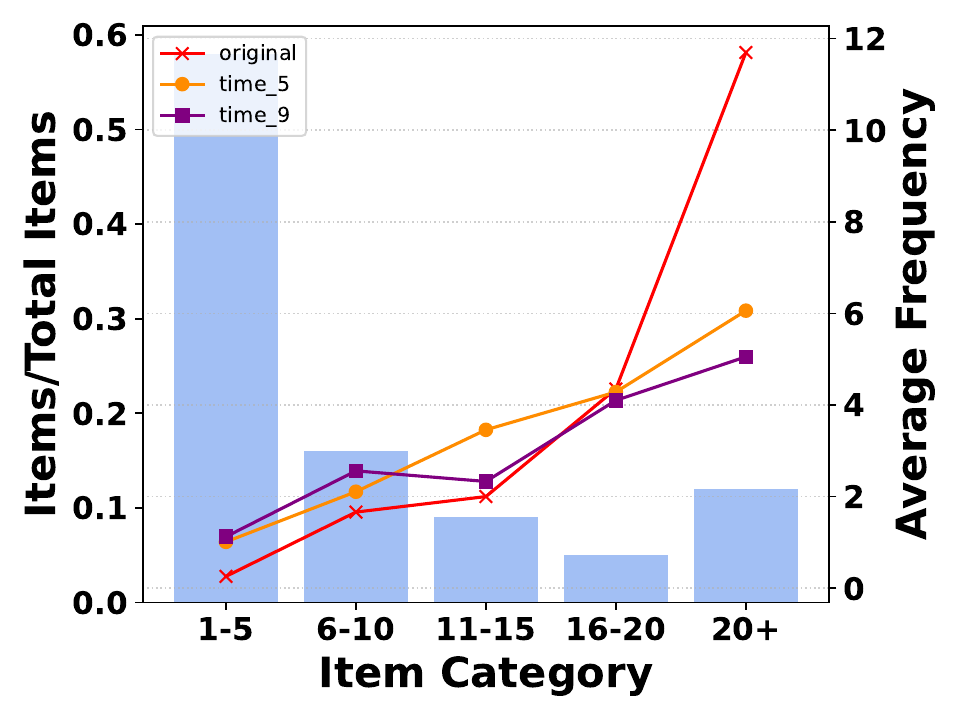}
        \caption{with $\alpha = 1$ and $\lambda = 5$ }
    \end{subfigure}
    \hfill 
    \begin{subfigure}[b]{0.48\linewidth}
    \captionsetup{belowskip=0pt, aboveskip=0pt}
        \includegraphics[width=\linewidth]{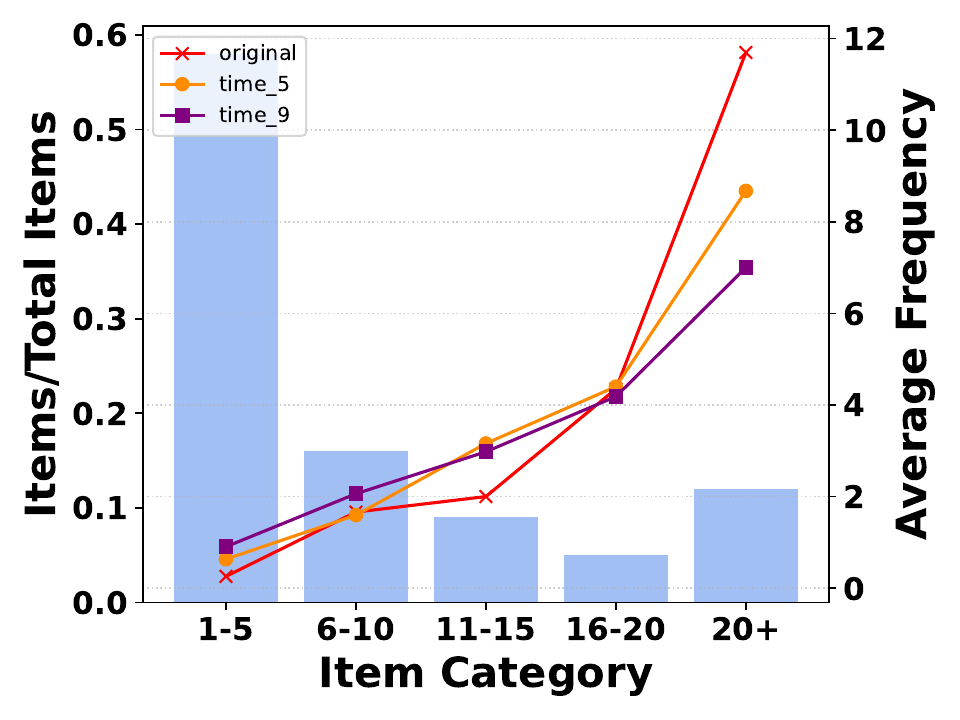}
        \caption{with $\alpha = 3$ and $\lambda = 5$}
    \end{subfigure}
    \hfill
    \begin{subfigure}[b]{0.48\linewidth}
    \captionsetup{belowskip=0pt, aboveskip=0pt}
        \includegraphics[width=\linewidth]{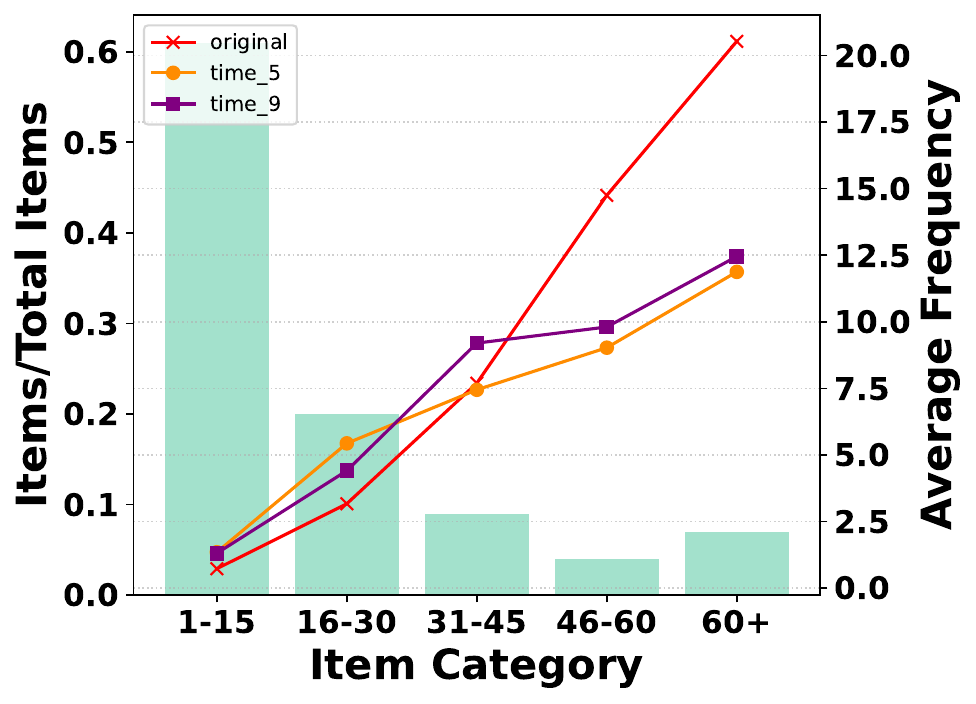}
        \caption{with $\alpha = 1$ and $\lambda = 2$ }
    \end{subfigure}
    \hfill 
    \begin{subfigure}[b]{0.48\linewidth}
    \captionsetup{belowskip=0pt, aboveskip=0pt}
        \includegraphics[width=\linewidth]{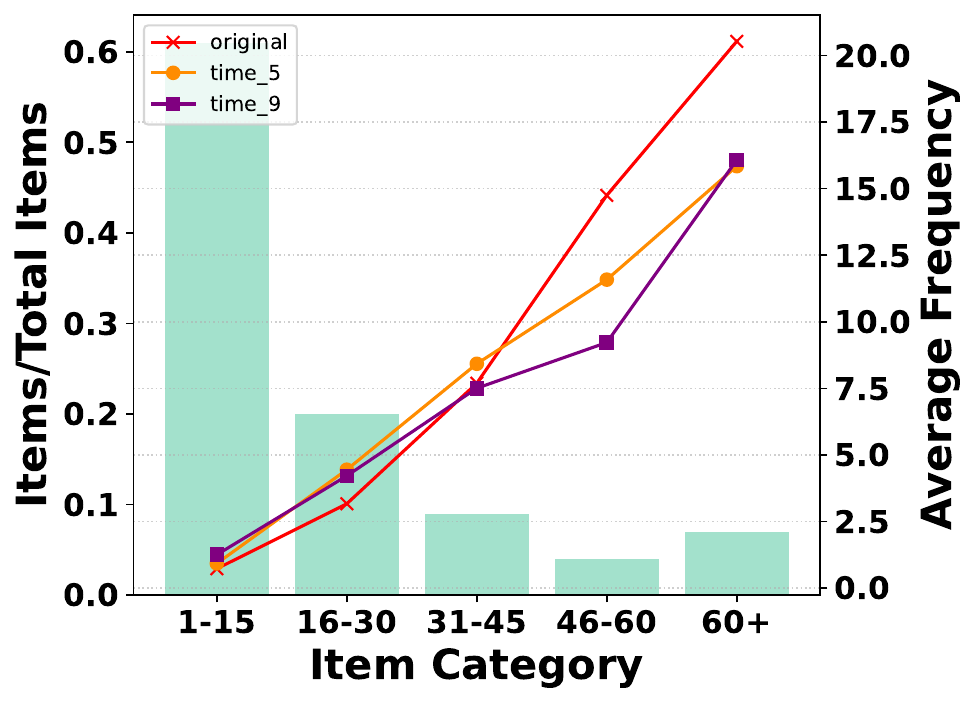}
        \caption{with $\alpha = 3$ and $\lambda = 2$}
    \end{subfigure}
    \caption{Average frequency of different item categories in two datasets recommended by strategic agents when $k=10$.}
    \label{debias_result}
\end{figure}

Figure \ref{debias_result} illustrates the effectiveness of our proposed debiasing mechanism through the deployment of a re-ranking method driven by strategic agents, as evidenced in the top-$k$ recommendation lists in two datasets. The top two represent Yelp, while the bottom two represent the industrial.

The analysis of this subsection delves into the average frequency of recommendations across five distinct item categories, ordered by ascending popularity levels, at the fifth and ninth time step. Compared to original top-$k$ lists generated by $\mathcal{S}$ initially, our findings indicate a strategic reduction in the recommendation frequencies of highly popular items, juxtaposed with an enhancement in the visibility of tail items within user candidate lists. Meanwhile, the  debiasing degree can be flexibly adjusted based on different parameter settings. Notably, it reveals that employing a lower $\alpha$ value facilitates a rapid increase in the exposure of items at the lower end of the popularity spectrum, e.g. the category of [1-5,6-10] in Yelp and the category of [1-15,16-30] in the industrial. Conversely, a higher $\alpha$ value ensures that the suppression of popular items is moderated, e.g. the category of [20+] and [60+], preventing their excessive demotion in the recommendation process.

\section{Conclusion}
In our study, we delve into the issue of popularity bias within recommendation systems, from the perspective of producer strategic behavior in two-sided markets. Enabling to involve the producers in combating biases actively, our proposed approach proposed a re-ranking method driven by strategic agents, which can enrich the connotation of fair-exposure from the perspective of item features combination. In order to discover the beneficial features of tail items and make full use of them, we incentivize content creators to modify item features based on performative optimization results per round, thereby enhancing exposure for long-tail items efficiently while maintaining recommendation accuracy. The method employs the differentiable ranking operator, achieving dual optimization goals of accuracy and fair exposure in the end-to-end training paradigm. Experimental results on two real-world datasets validate the effectiveness and potential of our method for addressing long-tail item exposure in the long run.

\begin{acks}
This study was partially funded by the supports of National Natural Science Foundation of China (72101176).
\end{acks}

\bibliographystyle{ACM-Reference-Format}
\balance
\bibliography{sample-base}

\clearpage
\section*{Appendix}
\appendix
\section{Proof}
\label{PROOF}
We get closed-form expression for the best response of the strategic agent using Karush-Kuhn-Tucker (KKT) conditions. The detailed proof process is as follows. Firstly, the constrained optimization problem is given below:
$$
\max_{x'} w^T x' - \alpha \|x' - x\|_2^2 \quad \text{s.t.} \quad \|x'\|_2 = 1
$$
Next, we construct the Lagrangian \( L \) with the constraint incorporated by the Lagrange multiplier \( \gamma \):
$$
L(x', \gamma) = w^T x' - \alpha \|x' - x\|_2^2 + \gamma (\|x'\|_2^2 - 1).
$$
Take the gradient of \( L \) with respect to \( x' \) and set it to zero for optimality. The equation is as below:
$$
\nabla_{x'} L = w - 2\alpha(x' - x) + 2\gamma x' = 0,
$$ 
and because we need to find the optimal response form for $x'$, we then solve the equation for \( x' \) which gives:
$$
x' = \frac{w + 2\alpha x}{2(\alpha - \gamma)},
$$
and substituting \( x' \) into the constraint \( \|x'\|_2 = 1 \) gives us a new expression, which states the Euclidean norm of $x'$ is equal to 1:
$$
\left\| \frac{w + 2\alpha x}{2(\alpha - \gamma)} \right\|_2 = 1.
$$
This condition allows us to solve for \( \gamma \). Once \( \gamma \) is found, we can substitute back into the expression for \( x' \) to obtain the final expression that satisfies the constraint.
The final expression for \( x' \), taking into account the constraint, can be written as:
$$
x' = \frac{w + 2\alpha x}{\|w + 2\alpha x\|_2}. \quad
$$

\section{Additional Description}
\label{feature description}
In order to utilize the feature incentives of strategy agents, we collect as much semantic information as possible about items. For Yelp data, we used data processed by \cite{eilat2023performative}. Category data is leveraged to create supplementary features by clustering similar categories that share comparable contextual significance. For industrial, we collect from a large online mall. The complete features are as follows.

\textbf{Yelp restaurants: } `stars’, `alcohol’, `restaurants good for groups’, `restaurants reservations’, `restaurants attire’, `bike parking’, `restaurants price range’, `has tv’, `noise level’, `restaurants take out’, `caters’, `outdoor seating’, `good for meal-dessert’, `good for meal-late night’, `good for meal-lunch’, `good for meal-dinner’, `good for meal-brunch’, `good for meal-breakfast’, `dogs allowed’, `restaurants delivery’, `japanese’, `chinese’, `india’, `middle east’, `mexican food’, `sweets’, `coffee’, `italian’, `burgers’, `hot dogs’, `sandwiches’, `steak’, `pizza’, `seafood’, `fast food’, `vegan’, `ice cream’, `restaurants table service’, `business accepts credit cards’, `wheel chair accessible’, `drive thru’, `happy hour’, `corkage’.

\textbf{Laptops in the mall:  } `office laptop', `gaming laptop', `brand1', `brand2', `brand3', `brand4', `brand5', `brand6', `brand7', `brand8', `relative price', `color-black', `color-gray \& color-silver', `color-others', `screen size', `cpu-category1', `cpu-category2', `ram level', `disk level', `integrated video-card', `dedicated video-card', `portable', `installment', `private customization'.

Then, we undertake feature processing for training simulators. For the categorical attributes of items, we apply one-hot encoding and for numerical attributes, we implement normalization techniques. Additionally, for features possessing several levels, we employ a graded encoding strategy, utilizing numerical values ranging from 0 to 1 to represent the hierarchy levels.

\end{document}